\documentstyle[prd,aps]{revtex}
\begin{document}
\draft

\title{Factorial Moments in a Generalized Lattice Gas Model}

\author{T. Wettig and A. D. Jackson}

\address{NORDITA, Blegdamsvej 17, DK--2100 Copenhagen \O, Denmark and\\
Department of Physics, State University of New York, Stony Brook,
NY 11794-3800}

\date{\today}

\maketitle

\begin{abstract}
We construct a simple multicomponent lattice gas model in one
dimension in which each site can either be empty or occupied by at
most one particle of any one of $D$ species. Particles interact with a
nearest neighbor interaction which depends on the species involved.
This model is capable of reproducing the relations between factorial
moments observed in high--energy scattering experiments for moderate
values of $D$.  The factorial moments of the negative binomial
distribution can be obtained exactly in the limit as $D$ becomes
large, and two suitable prescriptions involving randomly drawn nearest
neighbor interactions are given.  These results indicate the need for
considerable care in any attempt to extract information regarding
possible critical phenomena from empirical factorial moments.
\end{abstract}


\section{Introduction}
\label{sect1}

Factorial moments provide a useful tool in the analysis of high energy
scattering data such as obtained in $e^+ e^-$ scattering \cite{ALEPH},
$p{\bar p}$ scattering (at energies up to 900 GeV) \cite{UA5} or the
scattering of protons or heavy ions ({\em e.g.,} $^{16}$O and
$^{32}$S) by heavy nuclei at a projectile energy of 200 GeV/A
\cite{KLM}.  One considers the full range of some variable (usually
the rapidity) for which one knows the average multiplicity, $\langle n
\rangle$, and its dispersion, $\langle \Delta n^2 \rangle = \langle
n^2 \rangle - \langle n \rangle^2$.  The data is then broken into $M$
equal bins, and one constructs the factorial moments as a suitably
normalized average of a combination of moments in each individual bin.
Specifically,
\begin{equation}
F_q (M) = \left[\frac{1}{M} \sum_{i=1}^M \langle n (n-1) (n-2) \ldots
(n-q+1) \rangle_i \right] \left/ \left[ \frac{1}{M} \sum_{i=1}^M
\langle n \rangle_i \right]^q \right. \ \ .
\label{1.1}
\end{equation}
Although interest in these forms has been motivated by theoretical
considerations \cite{Bialas1}, factorial moments are merely one way to
arrange the experimental data.  Evidently, $F_q (M)$ probes certain
combinations of the properties of the 1--, 2--, 3--, \ldots, $q$--body
correlations between particles in each bin.  If non--interacting
particles are distributed in the various bins in a purely statistical
fashion, all factorial moments are $1$ for all $M$.  This result is
completely inconsistent with the data.  Consider, for example, $p{\bar
p}$ scattering at 900 GeV.  Here, $F_2 (M)$ is seen to grow from 1
(for small $M$) to roughly 1.7 (for the largest values of $M$
considered).  Over the same range, $F_3 (M)$ ranges from 1 to 4.2,
$F_4 (M)$ ranges from 1 to 15 and $F_5 (M)$ ranges from 1 to 80.  This
growth in the factorial moments with $M$ is often called
`intermittency' in the literature and is frequently regarded as
indicating the presence of fluctuations of many different sizes.

It has been observed \cite{Giovannini} that the relations {\em
between\/} the various factorial moments for $p {\bar p}$ scattering
(and all of the other physical processes mentioned above) can be
reproduced with remarkable accuracy by the `negative binomial
distribution' (NB) for which
\begin{equation}
F_q^{NB} (M) = (1 + cM)(1 + 2cM) \ldots (1+[q-1]cM)
\label{1.2}
\end{equation}
with
\begin{equation}
c = \frac{\langle \Delta n^2 \rangle - \langle n \rangle}{\langle n
\rangle^2} \ \ .
\label{1.3}
\end{equation}
Plots of $F_q (M)$ versus $M$ evidently depend on the global averages
$\langle n \rangle$ and $\langle \Delta n^2 \rangle$.  Such plots will
be different for the various physical processes considered and are not
reproduced by the negative binomial distributions as given by
Eq.~(\ref{1.2}) \cite{footnote1}. However, since $F_2^{NB}$ is simply
$(1 + cM)$, it is tempting to consider the more `universal' plots of
$F_q$ versus $F_2$ \cite{Carruthers1}.  Such plots no longer depend on
the parameter $c$ and invite the comparison of data from very
different processes \cite{footnote2}. Such plots have been made over
the available range $1 < F_2 < 1.7$.  They do reveal universal
behaviour \cite{Ochs} and are in striking agreement with the curves
obtained from the negative binomial distribution.

The dramatic growth of the factorial moments with $M$ led Bialas and
Peschanski to note that `an observation of a variation in $\langle F_i
\rangle$ (our $F_q$) with $\delta y$ (our $M$) indicates the presence
of genuine fluctuations which must have some physical origin'
\cite{Bialas1}.  Many authors have taken up the challenge of
describing this physical origin \cite{Carruthers2}.  Some have
concentrated on the apparent presence of fluctuations of many
different sizes and considered models incorporating a variety of
critical phenomena \cite{Satz}.  Others have studied both schematic
and more realistic versions of cascade models
\cite{Bialas2,Giovannini}.

In two interesting papers, Chau and Huang have offered a different
kind of insight \cite{Chau}.  They imagine that the full range of
rapidity corresponds to the $N$ sites of a one--dimensional Ising (or
lattice gas) model \cite{footnote3}.  This model is exactly solvable.
They consider the $q$--body correlations and the factorial moments
which come from a lattice gas model (in the limit $N \rightarrow
\infty$).  The two parameters of the Hamiltonian are determined by
fixing the global values of $\langle n \rangle$ and $\langle \Delta
n^2 \rangle$.  The resulting factorial moments have the form
\begin{equation}
F_q^{LG} (M) = \sum_{k=0}^{q-1} \frac{q!(q-1)!}{(q-k)!k!(q-1-k)! 2^k}
(cM)^k
\label{1.4}
\end{equation}
where $c$ is given by Eq.~(\ref{1.3}).  This result is both rather
more and rather less than meets the eye.

Let us address the `less' first.  While the lattice gas factorial
moments are not identical to the corresponding moments of the
negative binomial distribution, both share the common
small $M$ expansion
\begin{equation}
F_q (M) = 1 + \frac{q(q-1)}{2} cM + {\cal O}(c^2 M^2 ) \ \ .
\label{1.5}
\end{equation}
We have already noted that the leading term is due to `one--body
effects'.  It should come as no surprise that the term of order $M$
is precisely due to two--body correlations.  Under the assumption that
genuine two--body correlations have a finite range, the $q$--dependence
of the term of order $M$ is uniquely determined.  The coefficient $c$,
which can be expressed as a suitable integral of the two--body
correlation function, is also fixed by the global dispersion, $\langle
\Delta n^2 \rangle$.  Thus, the observation of Chau
and Huang that the factorial moments `given by single negative
binomials are almost exactly the same as ours for $F_2 \rightarrow
1$' is a trivial consequence of the rules of the game ({\em i.e.,}
the fixing of $\langle n \rangle$ and $\langle \Delta n^2 \rangle$)
and the fact that the Ising model predicts many--body correlations of
a finite range.

Chau and Huang further note that the negative binomial distributions
are more successful than the lattice gas model for values of $F_2$
larger than 1.5.  This is the domain where terms of order $M^2$ and
higher --- which are not model independent --- play a role.  A few
comments on the existing data are in order before continuing.  Data
for 200 GeV/A $^{16}$O and $^{32}$S scattering on emulsions is limited
to the range $1 < F_2 < 1.35$.  Thus, this data never really probes
the interesting (model dependent) regions of $F_2$.  (Over this range,
the difference between $F_3^{LG}$ and $F_3^{NB}$ is less than 2.8\%
which is small compared with the experimental uncertainties.)  The
data for 200 GeV $p$ on emulsion and for 900 GeV $p{\bar p}$
scattering has a somewhat wider range covering $1 < F_2 \le 1.7$.  At
the upper end of this range, the difference between lattice gas and
negative binomial forms for $F_3$ grows to 6.2\% while the non--linear
terms in the negative binomial form account for 24\% of $F_3$.
Uncertainties in the data in this region (both systematic and
statistical) are approximately 7.2\%.  The situation is similar for the
existing data on $F_4$ and $F_5$.  It is possible to find convincing
empirical evidence for higher than linear terms in $F_q$.  It is more
difficult to distinguish empirically between negative binomial and
lattice gas models although the former does provide a better global
fit.

We believe that Chau and Huang have provided something very positive
in their work.  Their model admits an elementary generalization which
shows one possible route which leads smoothly from the lattice gas
model to the negative binomial distributions \cite{JWB}.  While the
nature of this route may be of limited empirical value given the
current status of the data, it strikes us as being of some theoretical
importance.  This generalization, which is the primary focus of this
paper, is simply stated: Consider a one dimensional lattice gas model
(again in the limit where the number of sites, $N$, is infinite) where
each site can either be empty or occupied by one particle which can be
of any one of $D$ species.  Each species has a chemical potential,
$\mu_d$, and each pair of species has a nearest neighbor interaction
of strength $\epsilon_{dd'}$.  Like the ordinary lattice gas (with $D
= 1$), this model allows for the analytic determination of the
\mbox{$q$--body} correlation functions and the factorial moments,
$F_q (M)$, using the obvious generalization of standard techniques.
As usual, this involves the construction and diagonalization of a
matrix, ${\cal M}$, related to the partition function for one pair of
adjacent sites.

The factorial moments for this generalized lattice gas model can then
be specified exactly in terms of certain (weighted) moments of an
elementary function of the eigenvalues of ${\cal M}$.  We shall show
that, for any value of $D$, these moments can be chosen to reproduce
all factorial moments of the ordinary lattice gas model.  Further, we
shall show that these moments can also be chosen to approach the
factorial moments of the negative binomial distribution.  In the limit
$D \rightarrow \infty$, this model permits the exact reproduction of
{\em all\/} factorial moments of the negative binomial distribution.

Given the specific form of ${\cal M}$, there is no guarantee that
these conditions on its eigenvalues can be met for any choice of the
parameters appearing in the related Hamiltonian.  We shall, thus, give
two methods for the selection of the various chemical potentials and
nearest neighbor interaction parameters in ${\cal M}$ which explicitly
reproduce the factorial moments of the negative binomial distribution.
This choice is essentially a `random dynamics' model in which the $D$
chemical potentials are set equal and in which the various nearest
neighbor interactions are drawn at random according to a certain
distribution.  The random nature of this model and the smallness of
the dispersion in its factorial moments for $D$ finite and small
provides some understanding for the success of cascade model
calculations of these processes and for the relative insensitivity of
such calculations to the details of input parameters.

The organization of this paper will be as follows.  In
Sec.~\ref{sect2} we summarize the results of Chau and Huang.  One
purpose of this summary is to establish the notation of the lattice
gas model (as opposed to the Ising model) which will later be
generalized.  In the process we shall establish a useful (and more
general) theorem regarding the factorial moments and shall explain
precisely which features of the Ising model are responsible for the
form of Eq.~(4).  Along the way, we shall point out why the general form of
Eq.~(5) is more a matter of definition than of physical content.

In Sec.~\ref{sect3} we shall establish our generalization to a
multicomponent lattice gas model containing $D$ species, outline the
techniques for its analytic solution, and present the general form for
the resulting factorial moments.

In Sec.~\ref{sect4} we shall consider the constraints which must be
imposed on this generalized model if it is (i) to reproduce the
results of Chau and Huang or (ii) to reproduce the factorial moments
of the negative binomial distribution.  These constraints will be
established analytically as a well--defined distribution of the
eigenvalues of a certain $(D+1)$--dimensional matrix.  We shall also
report the results of numerical studies which allow us to express
these constraints in terms of a well--defined but random distribution of
the $D(D+1)/2$ nearest--neighbor interaction strengths between the $D$
species.

A variety of conclusions will be drawn in Sec.~\ref{sect5}.

Two short appendices are provided to deal with more technical matters.

\section{The Lattice Gas Model}
\label{sect2}

Here, we follow the arguments of Chau and Huang with the notational
difference that we consider a one--dimensional lattice gas rather than
an Ising model.  We consider $N$ microscopic sites which can have an
occupation number of either 0 or 1.  Particles occupying adjacent
sites will experience a nearest neighbor interaction of strength
$\epsilon$.  The system is assumed to be cyclic in the sense that
particle $N$ interacts with particle $1$ as well as with particle $(N-1)$.
The Hamiltonian for the system is simply
\begin{equation}
H = \epsilon [n_1 n_2 + n_2 n_3 + \ldots + n_{N-1} n_N + n_N n_1 ] \ \ .
\label{2.1}
\end{equation}
The partition function for this system is \cite{footnote4}
\begin{equation}
\sum \exp{\left[-H - \mu \sum_{i=1}^N n_i \right] } \ \ .
\label{2.2}
\end{equation}
The external summation here covers the $2^N$ terms where each
$n_i$ has the value 0 or 1.  This problem is rendered trivial by
constructing a two--dimensional matrix, ${\cal M}$, such that
\begin{equation}
{\cal M}_{n_1 n_2} = \exp{\left[-\epsilon n_1 n_2 - \frac{1}{2} \mu n_1
- \frac{1}{2} \mu n_2 \right]} \ \ .
\label{2.3}
\end{equation}
It is convenient to let the matrix indices run over 0 and 1.

The partition function for the system is now simply the trace of
${\cal M}^N$.  This trace is most easily constructed by defining the
orthogonal matrix, $\theta$, which diagonalizes ${\cal M}$.
\begin{equation}
{{\cal M}_d} = \theta^T {\cal M} \theta \ \ .
\label{2.4}
\end{equation}
The diagonal form ${{\cal M}_d}$ has two eigenvalues, $\lambda_{\pm}$,
and we shall choose $\theta$ such that $({{{\cal M}_d}})_{00}$ is the
eigenvalue of larger magnitude, $\lambda_+$.  (Given the form of ${\cal
M}$, it is clear that $\lambda_+ > 0$.)  We then find that
\begin{equation}
{\rm Tr}[{\cal M}^N ] = {\rm Tr}[\theta {{\cal M}_d}^N \theta^T ] =
{\rm Tr}[{{\cal M}_d}^N ] = \lambda_+^N + \lambda_-^N \ \ .
\label{2.5}
\end{equation}
In the limit $N \rightarrow \infty$, we can neglect the term
$\lambda_-^N$ at the cost of introducing an error which is
exponentially small in $N$.

In order to calculate the various correlation functions, we introduce
the number operator matrix for site $i$, $n_i$.  Clearly,
$(n_i )_{n_1 n_2} = \delta_{1 n_1} \delta_{1 n_2}$, {\em i.e.,}
\begin{equation}
n_i = \left( \begin{array}{cc} 0 & 0 \\ 0 & 1 \end{array} \right) \ \ .
\label{2.6}
\end{equation}
The number operator at each site is an idempotent with $n_i^2 = n_i$.
Since all sites are equivalent (due to the cyclic form of
Eq.~(\ref{2.1})), the average number of particles per site is simply
\begin{equation}
\langle n_i \rangle = \left. {\rm Tr} \left[ n_i {\cal M}^N \right]
\right/ {\rm Tr} \left[ {\cal M}^N \right] \ \ .
\label{2.7}
\end{equation}
Given the fact that ${\rm Tr}[{\cal M}^N]=\lambda_+^N$ in the large
$N$ limit, we see that it is useful to replace ${{\cal M}_d}$ by
another diagonal matrix, ${{\bar {\cal M}_d}}$, in which each diagonal
element is simply divided by $\lambda_+$.  Thus, $({{\bar {\cal
M}_d}})_{00} = 1$ and $({{\bar {\cal M}_d}})_{11} = {{\bar
\lambda}_{-}}$ where $|{\bar \lambda}_{-}| < 1$.  With this notation,
\begin{equation}
\langle n_i \rangle = {\rm Tr} \left[ n_i \theta {{\bar {\cal M}_d}}^N
\theta^T \right] \ \ .
\label{2.8}
\end{equation}
Given the trivial structure of $n_i$,
\begin{equation}
\langle n_i \rangle = (\theta {{\bar {\cal M}_d}}^N \theta^T )_{11}
\rightarrow (\theta_{10})^2
\label{2.9}
\end{equation}
where the last form applies in the limit $N \rightarrow \infty$.
While $\theta_{10}$ and ${{\bar \lambda}_{-}}$ can be determined in
terms of $\epsilon$ and $\mu$ following an explicit diagonalization of
${\cal M}$, this is not necessary.  It is our intention to impose
global constraints on $\langle n \rangle = N \langle n_i \rangle$ and
$\langle \Delta n^2 \rangle$.  This can be done with equal ease
working with $\theta_{10}$ and ${{\bar \lambda}_{-}}$.

The two--body correlation function can now be written immediately as
\begin{equation}
\langle n_i n_{i+j} \rangle = (\theta {{\bar {\cal M}_d}}^j \theta^T )_{11}
(\theta {{\bar {\cal M}_d}}^{N-j} \theta^T )_{11} \ \ .
\label{2.10}
\end{equation}
Taking the large $N$ limit and making the usual assumption that $N-j$
is always ${\cal O}(N)$, we find
\begin{equation}
\langle n_i n_{i+j} \rangle = \langle n_i \rangle [ \langle n_i \rangle
+ (1 - \langle n_i \rangle ) {{\bar \lambda}_{-}}^j ] \ \ .
\label{2.11}
\end{equation}
Eq.~(\ref{2.11}) is valid for $j \ge 0$.  The structure of this
two--body correlation function is instructive and very general.  For
large separations (large $j$), $\langle n_i n_{i+j} \rangle$
approaches the uncorrelated, statistical value of $\langle n_i
\rangle^2$.  The approach to this asymptotic value is exponentially
fast.  For small $j$, there are significant short--range correlations.
For $j=0$, we find $\langle n_i n_{i+j} \rangle = \langle n_i \rangle$
independent of ${\bar \lambda}_{-}$ which is merely a
reflection of the fact that the number operator is idempotent.

It is elementary to determine the global dispersion, $\langle \Delta
n^2 \rangle$.  Since it is our intention to constrain $\langle \Delta
n^2 \rangle$ to be consistent with data, let us deal with this problem
immediately.  We first construct
\begin{equation}
\langle n^2 \rangle = \sum_{i_1 , i_2  = 1}^N \langle n_{i_1} n_{i_2}
\rangle = N \langle n_i \rangle + 2 \sum_{i_1 < i_2 } \langle n_{i_1}
n_{i_2} \rangle \ \ .
\label{2.12}
\end{equation}
It is, of course, possible to do the summations in Eq.~(\ref{2.12})
exactly given the form of Eq.~(\ref{2.11}).  For purposes of later
arguments, it is more useful to obtain $\langle n^2 \rangle$
approximately by making approximations which neglect terms of order
$1/N$.  Thus, we write
\begin{equation}
\langle n^2 \rangle = \langle n \rangle + 2 \langle n_i \rangle^2
\sum_{i_1 < i_2} 1 + 2\langle n_i \rangle (1 - \langle n_i \rangle)
\sum_{i_1 < i_2 } {{\bar \lambda}_{-}}^{i_2 - i_1} \ \ .
\label{2.13}
\end{equation}
The first sum in Eq.~(\ref{2.13}) will be approximated by $N^2 /2$.
The second sum is more interesting.  Since ${{\bar \lambda}_{-}}^j$
converges exponentially with $j$, we shall allow the $i_2$ sum to
extend over all values $1 \le (i_2 - i_1 ) \le \infty$.  The sum over
$i_1$ then merely introduces a factor of $N$.  Further, we neglect
$\langle n_i \rangle$ compared to $1$.  These approximations each
introduce errors of order $1/N$ which are acceptable as $N \rightarrow
\infty$.  We immediately obtain
\begin{equation}
\langle n^2 \rangle = \langle n \rangle + \langle n \rangle^2 +
2 \langle n \rangle \frac{{{\bar \lambda}_{-}}}{1 - {{\bar \lambda}_{-}}}
\ \ .
\label{2.14}
\end{equation}
We are now able to set ${{\bar \lambda}_{-}}$ in order to reproduce
any desired dispersion.

The calculation of factorial moments can be simplified significantly
when the number operator is idempotent.  This issue is addressed in
Appendix~\ref{app1} where appropriate operator (and ensemble average)
identities are established.  Using the definition of the factorial
moments, Eq.~(\ref{1.1}), using the simplifying result of
Eq.~(\ref{a.6}), and adopting the same large $N$ approximations, we
can obtain $F_2 (M)$.  (Now, the sums analogous to those in
Eq.~(\ref{2.12}) extend to an upper limit of $N_B$ where $N_B = N /
M$.  Since the number of bins, $M$, is finite, we are also concerned
with the limit $N_B \rightarrow \infty$.)  We find
\begin{equation}
F_2^{LG} (M) = 1 + c M
\label{2.15}
\end{equation}
where
\begin{equation}
c = \frac{2}{\langle n \rangle} \frac{{{\bar \lambda}_{-}}}{1 - {{\bar
\lambda}_{-}}} \ \ .
\label{2.16}
\end{equation}
Returning to Eqs.~(\ref{2.10}) and (\ref{2.11}), it is now elementary
to construct the $q$--body correlation functions as
\begin{equation}
\langle n_{i_1} n_{i_1 + i_2} \ldots n_{i_1 + \ldots + i_q} \rangle
= \langle n_i \rangle [\langle n_i \rangle + (1 - \langle n_i \rangle )
{{\bar \lambda}_{-}}^{i_2} ] [\langle n_i \rangle + (1-\langle n_i \rangle )
{{\bar \lambda}_{-}}^{i_3} ] \ldots [\langle n_i \rangle + (1 -
\langle n_i \rangle ) {{\bar \lambda}_{-}}^{i_q} ] \ \ .
\label{2.17}
\end{equation}
Eq.~(\ref{2.17}) applies for $i_1,i_2,\ldots,i_q \ge 0$.  This form,
which is not unique to the lattice gas model, determines the form of
the factorial moments and merits comment.  The $q$--body correlation
function is completely determined by the two--body correlation
function and $\langle n_i \rangle$.  More precisely, the $q$--body
correlation function is a product of the $(q -1 )$ two--body
correlators between adjacent particles.  Finally, the individual
two--body correlators are given as the sum of a statistical term and a
short--range piece.  These qualitative features are sufficient to
determine the associated factorial moments uniquely without any
detailed information about the precise nature of the short--range part
of the two--body correlator.  To emphasize this independence, we shall
make the replacement of $(1 - \langle n_i \rangle ) {{\bar
\lambda}_{-}}^j$ by $g(j)$ where the only restriction on $g(j)$ is
that it is of short range.

Now let us turn to an arbitrary factorial moment in the limit $N
\rightarrow \infty$.  We again use the operator identity derived in
Appendix~\ref{app1}, Eq.~(\ref{a.6}), and replace sums by integrals to
obtain
\begin{eqnarray}
F_q^{LG} (M) & = & q! \left( \frac{M}{\langle n \rangle} \right)^{q}
\int_0^{N_B} dx_1 \int_0^{N_B - x_1} dx_2 \ldots \int_0^{N_B - x_1 -
\ldots - x_{q-1}} dx_q \nonumber \\
& {} & \ \ \ \ \ \ \ \ \ \ \ \ \ \ \ \ \times \frac{\langle n \rangle}{N}
[\frac{\langle n \rangle}{N} + g (x_2 ) ] [\frac{\langle n \rangle}{N} +
g (x_3 ) ] \ldots [\frac{\langle n \rangle}{N} + g (x_q ) ] \ \ .
\label{2.18}
\end{eqnarray}
The structure of this result is clear.  We now imagine expanding the
$(q-1)$ square brackets.  Since the function $g(x_i )$ is short range,
we can extend the $x_i$ integration to infinity for any of those
$2^{q-1}$ terms which contains a factor of $g(x_i )$.  Let us
introduce the notation
\begin{equation}
G = \int_0^{\infty} \ dx \ g(x) \ \ .
\label{2.19}
\end{equation}
The term ${\cal O}(M^0 )$ can only come from picking the statistical
factor in each term.  There is $1$ way to do this.  We obtain no
factors of $G$.  The remaining integrals give a factor of $N_B^q /q!$.
Hence, the leading term in $F_q^{LG} (M)$ is always one, as expected.
The term ${\cal O}(M)$ comes from picking $(q-1)$ statistical factors
and $1$ factor of $g$.  There are $(q-1)$ ways to do this, and we
obtain a factor of $G$.  The remaining integrals give a factor of
$N_B^{q-1}/(q-1)!$.  Hence, the next term in $F_q^{LG} (M)$ is
precisely $(2GM / \langle n \rangle) q(q-1)/2$.  The general result is
now obvious:
\begin{equation}
F_q^{LG} (M) = \sum_{k=0}^{q-1} \left[ \frac{2 G M}{\langle n \rangle}
\right]^k
\frac{q! (q-1)!}{(q-k)!k!(q-1-k)! 2^k } \ \ .
\label{2.20}
\end{equation}
Comparing Eq.~(\ref{2.20}) with $q=2$ with the result of
Eq.~(\ref{2.15}), we see that we can make the identification
\begin{equation}
c = \frac{2G}{\langle n \rangle} \ \ .
\label{2.21}
\end{equation}

Eqs.~(\ref{2.20}) and (\ref{2.21}) are precisely the forms found by
Chau and Huang as quoted in Eq.~(\ref{1.4}).  The present derivation
makes it clear that these results {\em do\/} depend on the fact that
the $q$--body correlator is a product of the $(q-1)$ consecutive
two--body correlators and that they do {\em not\/} depend on the
specific form of the short--range two--body correlations.

The fact that
\begin{equation}
F_q (M) = 1 + \frac{q(q-1)}{2} cM + {\cal O}(c^2 M^2 )
\label{2.22}
\end{equation}
is even more general.  It depends only on the fact that the $q$--body
correlation function contains no long--range terms and thus includes
all one--dimensional models.  For any model we can consider
\begin{equation}
\langle n_{i_1} n_{i_1 + i_2} \ldots n_{i_1 + \ldots + i_q}  \rangle
\label{2.23}
\end{equation}
in the limit where one of the indices ({\em e.g.}, $i_k$) is small and
all others are large.  In the absence of long--range correlations,
this limit must reduce to
\begin{equation}
\langle n_i \rangle^{q-2} \langle n_i n_{i + i_k }  \rangle \ \ .
\label{2.24}
\end{equation}
(Translational invariance ensures that the final term here depends
only on $i_k$.)  Such terms are the only ones which contribute to the
linear piece of $F_q (M)$.  Their $q$ dependence is protected by
elementary combinatorics.  The coefficient $c$ is protected by the
fact that the two--body correlation has been adjusted to fit the
empirical global dispersion, $\langle \Delta n^2 \rangle$.  In a
similar fashion, the terms ${\cal O}(M^2 )$ receive contributions from
both two-- and three--body correlations.  If we had insisted on
establishing global constraints on $\langle n^3 \rangle$, these terms
would be similarly protected.

Thus, the fact that lattice gas factorial moments agree with those
coming from the negative binomial distribution in the limit $cM
\rightarrow 0$ provides {\em no\/} indication that the lattice gas
model is correct.  It is merely an indication of the fact that there
are no long--range correlations and that the `rules of the game' are
to consider only models with fixed global $\langle n \rangle$ and
$\langle \Delta n^2 \rangle$.

\section{A Generalized Lattice Gas Model}
\label{sect3}

As noted in Sec.~\ref{sect2}, the ordinary lattice gas model does
not reproduce the factorial moments of the negative binomial
distribution.  The fact that there is agreement in the small $cM$
limit is a consequence of general properties of one--dimensional
models and is not indicative of any particular merit of the lattice
gas model.  In this section we shall construct a simple generalization
of this lattice gas model which has the capacity to reproduce the
negative binomial factorial moments {\em exactly}.  This model is a
conceptually simple one--dimensional model which is also exactly
solvable (in a sense which is appropriate for constructing factorial
moments).

Consider a lattice gas in which site $i$ is either unoccupied or is
occupied by at most one particle which can be of any one of $D$
species.  Each species has its own chemical potential, $\mu_d$.
Particles again have a nearest neighbor interaction.  The strength of
the interaction between a particle of species $d$ and a particle of
species $d'$ is $\epsilon_{d d'}$.

This model can be solved using precisely the standard techniques of
Sec.~\ref{sect2}.  The matrix ${\cal M}$ now becomes a real,
symmetric matrix of dimension $(D+1)$.  The elements in this matrix
are
\begin{eqnarray}
{\cal M}_{00} & = & 1 \nonumber \\
{\cal M}_{0d} & = & \exp{[-\mu_d / 2]} \label{3.1} \\
{\cal M}_{dd'} & = & \exp{[-\epsilon_{dd'} -\mu_d / 2 - \mu_{d'} / 2]}
\nonumber
\end{eqnarray}
(We allow the matrix indices to run from $0$ to $D$. Again, the
inverse temperature, $\beta$, has been set equal to $1$.)  The number
operator at site $i$ is also a $(D+1)$--dimensional matrix having the
form $n_i = \openone - T$ with $T_{dd'} = \delta_{d0}
\delta_{d'0}$, {\em i.e.,}
\begin{equation}
n_i = \left( \begin{array}{cccccc}
                0 & & & & & \\
                & 1 & & & & \\
                & & 1 & & & \\
                & & & 1 & & \\
                & & & & \ddots & \\
                & & & & & 1
             \end{array} \right) \ \ .
\label{3.2}
\end{equation}
Both this number operator and $T$ are idempotent, so that the results
of Eqs.~(\ref{a.3}) and (\ref{a.6}) remain valid both at the operator
level and at the level of ensemble averages.

We again find the transformation $\theta$ which diagonalizes ${\cal
M}$ and places the eigenvalue of largest magnitude in $({\cal
M}_d)_{00}$ \cite{footnote5}. As before, we construct the
matrix ${\bar {\cal M}}_d$ by dividing each element of ${\cal M}_d$ by
the eigenvalue of largest magnitude.  (The form of ${\cal M}$ again
ensures that this largest eigenvalue will be positive.)  Thus,
$({\bar{\cal M}}_d)_{00}$ is again $1$. While we shall retain this
notation, it is not completely necessary in practice. The spirit of
the model is, ultimately, to take the limit as $N \rightarrow \infty$
for fixed $\langle n \rangle$.  In this limit, the average occupation
per site, $\langle n_i \rangle$, tends to zero.  This limit will be
realized by taking the limit as all of the $\mu_d \rightarrow
+\infty$.  In this limit, the coupling of the $D \times D$ submatrix
(with $d,d' \ne 0$) to the remaining elements of ${\cal M}$ becomes
arbitrarily weak.  In the limit, it is legitimate to treat this
coupling in lowest--order perturbation theory. We shall return to this
point in Sec.~\ref{sect4}.

With these preliminaries in hand, the construction of the various
correlation functions and factorial moments proceeds as before.  The
average number of particles per site is simply
\begin{equation}
\langle n_i  \rangle = \frac{\langle n  \rangle}{N} = Tr[(\openone - T)
\theta {{\bar {\cal M}_d}}^N \theta^T ]
\label{3.4}
\end{equation}
In the limit $N \rightarrow \infty$, ${{\bar {\cal M}_d}}^N$ may be
set equal to $T$ with exponentially small errors.  The two terms in
the factor $(\openone -T)$ must be treated separately
\cite{footnote6}. One immediately finds
\begin{equation}
\langle n_i  \rangle = 1 - (\theta_{00})^2
\label{3.5}
\end{equation}
As usual, $\langle n \rangle$ and hence $(\theta_{00})^2$ will be
fixed by experiment.

The two--body correlator is readily calculated as
\begin{equation}
\langle n_i n_{i+j}  \rangle = Tr[(\openone - T) \theta
{{\bar {\cal M}_d}}^j \theta^T (\openone - T) \theta T \theta^T ] \ \ .
\label{3.6}
\end{equation}
Expanding the factors of $(\openone -T)$, one immediately obtains
\begin{equation}
\langle n_i n_{i+j}  \rangle = \langle n_i  \rangle [ \langle n_i  \rangle
+ (1 - \langle n_i  \rangle ) \sum_{\ell=1}^D
(\theta_{0 \ell})^2 {{\bar \lambda}_{\ell}}^j ] \ \ .
\label{3.7}
\end{equation}
This form is virtually identical to that of Eq.~(\ref{2.11}).  The
first term is again the asymptotic statistical probability of finding one
particle (of any species) on each of the two sites.  Since $|{\bar
\lambda}_{\ell}| < 1$, the second term here again vanishes exponentially
for large $j$.  However, in this case, we are confronted with a sum of
exponentials rather than a single term.

Following the arguments from Eq.~(\ref{2.12}) to (\ref{2.16}) we find
that \cite{footnote7}
\begin{equation}
F_2^{GLG} (M) = 1 + cM
\label{3.8}
\end{equation}
with
\begin{equation}
c = \frac{\langle n^2 \rangle - \langle n \rangle^2 - \langle n \rangle}
{\langle n \rangle^2} = \frac{2}{\langle n \rangle} \sum_{\ell=1}^D
b_\ell^2 \frac{{\bar \lambda}_\ell}{1 - {\bar \lambda}_\ell} \ \ .
\label{3.9}
\end{equation}
Here, we have taken the notational liberty of exploiting the fact that
$\theta$ is an orthogonal matrix and made the substitution
\begin{equation}
b_\ell^2 = \frac{(\theta_{0\ell})^2}{1-\langle n_i \rangle} \ \ \ \
{\rm with} \ \ \ \ \sum_{\ell=1}^D b_\ell^2 = 1 \ \ .
\label{3.10}
\end{equation}
Again, the similarities between Eqs.~(\ref{3.9}) and (\ref{2.16}) are
striking.  The only difference is the presence of the sum over $\ell$.

Since certain differences arise at the level of the three--body
correlation function, this term is worth discussing specifically.
Obviously,
\begin{equation}
\langle n_i n_{i+j} n_{i+j+k}  \rangle = Tr[(\openone -T) \theta {{\bar
{\cal M}_d}}^j \theta^T (\openone -T) \theta {{\bar {\cal M}_d}}^k
\theta^T (\openone -T) \theta T \theta^T ]
\label{3.11}
\end{equation}
Expanding the factors of $(\openone -T)$ and constructing the traces
gives rise to the form
\begin{eqnarray}
\langle n_i n_{i+j} n_{i+j+k}  \rangle & = & \langle n_i  \rangle^3 +
\langle n_i  \rangle^2 (1 - \langle n_i  \rangle )
[\sum b_{\ell}^2 {{\bar \lambda}_{\ell}}^j +
\sum b_{\ell}^2 {{\bar \lambda}_{\ell}}^k -
\sum b_{\ell}^2 {{\bar \lambda}_{\ell}}^j \sum b_{\ell}^2
{{\bar \lambda}_{\ell}}^k ] \nonumber \\
& {} & + \langle n_i  \rangle (1 - \langle n_i  \rangle) \sum b_{\ell}^2
{{\bar \lambda}_{\ell}}^{j+k}
\label{3.12}
\end{eqnarray}
This result differs from the expression for the ordinary lattice gas.
The three--body correlator {\em cannot} be expressed as a product of
consecutive two--body correlators.  This is important since it will
give us precisely the freedom we need to proceed from the lattice gas
factorial moments to those of the negative binomial distribution.
Nevertheless, this correlator is still determined by the two--body
correlation function.  It merely requires the addition of $\langle n_i
n_{i+j+k} \rangle$.  This fact is also important, since we seek
factorial moments which depend on the global properties of the
distribution through $c$ only.  Note that in the special case of $D=1$
appropriate for the ordinary lattice gas, the sums collapse to a
single term with $b_1^2 = 1$.

It is now straightforward to perform the sums necessary to determine
$F_3^{GLG} (M)$ in the limit as $N \rightarrow \infty$.  Using
Eq.~(\ref{a.6}) we find
\begin{equation}
F_3^{GLG} (M) = 1 + 3cM + \frac{3}{2} d^2 M^2 \ \ .
\label{3.13}
\end{equation}
The presence of the term involving $\langle n_i n_{i+j+k} \rangle$ in
the three--body correlation function has forced the introduction of
the new term
\begin{equation}
d^2 = \frac{4}{\langle n \rangle^2} \sum_{\ell=1}^D b_{\ell}^2 \left(
\frac{{\bar \lambda}_{\ell}}{1 - {\bar \lambda}_{\ell}} \right)^2
\label{3.14}
\end{equation}
We note that this third factorial moment will equal that of the
original lattice gas model provided that $d^2 = c^2$.  There are (at
least) two situations where this will happen.  One is the case where
precisely one of the $b_{\ell}^2$ is equal to $1$ (with the remaining
$b_{\ell}^2$ equal to $0$).  The other is the case where all of the
${\bar \lambda}_{\ell}$ are equal.

We now turn to the general case of $F_q^{GLG} (M)$.  The strategy is
now perfectly straightforward and exceptionally tedious.  We construct
the $q$--body correlation function as the obvious generalization of
Eqs.~(\ref{3.6}) and (\ref{3.11}).  We expand the various factors of
$(\openone -T)$ and perform the requisite traces.  Using
Eq.~(\ref{a.6}), we then perform the requisite bin sums.  The final
result follows from some algebra and combinatorics.  We find
\begin{equation}
F_q^{GLG} (M) = \sum_{k=0}^{q-1} f_k^{(q)} M^k
\label{3.15}
\end{equation}
where
\begin{equation}
f_k^{(q)} = \frac{q!}{k! (q-k)!} \left[ \frac{d^k}{dz^k} [ 1 +
\sum_{\mu=1}^{\infty} z^{\mu} s_{\mu}] ^{(q-k)} \right]_{z=0} \ \ .
\label{3.16}
\end{equation}
Here, we have introduced the definition
\begin{equation}
s_{\mu} = \frac{1}{\langle n \rangle^{\mu}} \sum_{\ell=1}^D
{b_{\ell}}^2 \left( \frac{{{\bar \lambda}}_{\ell}}{1-{{\bar
\lambda}}_{\ell}} \right)^{\mu} \ \
\label{3.17}
\end{equation}
which implies $s_1 = c/2$.

It is also desirable to derive a generating function from which the
factorial moments can be obtained by simple differentiation.  To this
end, we first observe that $f_k^{(q)}$ can be re--written as
\begin{equation}
f_k^{(q)} = \frac{1}{(q-k)!} \left[ \frac{d^q}{dz^q} z^{q-k}
[ 1 + \sum_{\mu=1}^{\infty} z^{\mu} s_{\mu} ] ^{(q-k)} \right]_{z=0} \ \ .
\label{3.18}
\end{equation}
We now change the summation index in Eq.~(\ref{3.15}) from $k$ to
$(q-k)$ to obtain
\begin{equation}
F_q^{GLG}(M) = M^q \left[ \frac{d^q}{dz^q} \sum_{k=1}^{q}
\frac{1}{k!} \left( \frac{z}{M} [ 1+\sum_{\mu=1}^{\infty}
s_{\mu} z^{\mu}] \right) ^k \right]_{z=0} \ \ .
\label{3.19}
\end{equation}
The sum over $k$ can be extended to range from $0$ to $\infty$ since
the corresponding derivatives are zero. The factorial moments thus
become
\begin{equation}
F_q^{GLG}(M) = M^q \left[ \frac{d^q}{dz^q} \exp{\left( \frac{z}{M}
[1+\sum_{\mu=1}^{\infty} s_{\mu} z^{\mu}] \right)} \right]_{z=0} \ \ .
\label{3.20}
\end{equation}
Eqs.~(\ref{3.15})--(\ref{3.17}) and (\ref{3.20}) represent the final
results of our generalized lattice gas model.

The various terms, $f_k^{(q)}$, which characterize the factorial
moments of our generalized lattice gas model are completely determined
by the leading terms $f_{k'}^{(k'+1)}$ for $k' \le k$.  This follows
from combinatoric arguments and has virtually nothing to do with the
underlying microscopic details of the model.  It does involve two
basic ingredients of the model.  First, that each site has an
occupancy of either $0$ or $1$.  Second, that the non--statistical
pieces of the various correlations are of short range
\cite{footnote8}. It was not {\em a priori\/} obvious that any
one--dimensional model would yield the factorial moments of the
negative binomial distribution. It is now clear, however, that the
generalized lattice gas model is capable of reproducing the factorial
moments of {\em any\/} distribution (including the NB) through a
suitable choice of the $s_\mu$. It remains to be demonstrated that
there exists a choice of the underlying Hamiltonian which will provide
the desired $s_\mu$. Such demonstration is the topic of the next
section.

\section{Obtaining the Factorial Moments of the Negative Binomial
Distribution}
\label{sect4}

As we have repeatedly emphasized, we are certainly able to reproduce
the factorial moments of the ordinary lattice gas model. We will show
in Appendix~\ref{app2} that this happens in the special case where
\begin{equation}
s_{\mu} = \left( \frac{c}{2} \right)^{\mu} \ \ .
\label{3.21}
\end{equation}
As noted, this special case can be realized either when the sums over
$\ell$ contain only one term or when all the reduced eigenvalues are
equal.

We are also able to pick the various terms, $s_{\mu}$, in such a way
as to reduce Eqs.~(\ref{3.15})--(\ref{3.17}) to Eq.~(\ref{1.2})
exactly.  It is readily verified that the choice
\begin{equation}
s_{\mu} = \frac{c^{\mu}}{\mu+1}
\label{4.1}
\end{equation}
establishes the desired equality.  (The demonstration of this fact is
given in Appendix~\ref{app2}.)  With this choice, the factorial
moments of our generalized lattice gas model are rendered {\em
identical\/} to the factorial moments of the negative binomial
distribution.  This answer poses a question.  Is it possible to pick
the dimension $D$ (the number of species) and to find a nearest
neighbor Hamiltonian and its related matrix, ${\cal M}$, such that the
constraints of Eq.~(\ref{4.1}) are satisfied?  As we shall see, it is.

As noted in the previous section, we are ultimately interested in the
$N \rightarrow \infty$ limit.  We shall arbitrarily set all chemical
potentials equal to $\mu_{\rm o}$ and realize this limit by taking
$\mu_{\rm o} \rightarrow \infty$.  It is now sufficient to consider
the $D$--dimensional submatrix, $\tilde{\cal M}$, obtained by
neglecting the 0-th row and column of ${\cal M}$. Using first order
perturbation theory, the coupling of $\tilde{\cal M}$ to the remaining
elements of ${\cal M}$ can then be treated exactly.  This will result
in a slightly different expression for $s_\mu$.  We obtain
\begin{equation}
s_\mu = \frac{1}{\langle n \rangle ^\mu} \sum_{k=1}^D a_k^2 \left(
\frac{{\tilde{\lambda}}_k}{1-{\tilde{\lambda}}_k} \right) ^\mu
\label{4.7.1}
\end{equation}
where the ${\tilde \lambda}_k$ are the eigenvalues of ${\tilde {\cal
M}}$ \cite{footnote9} and the $a_k^2$ are normalized coefficients
which follow from the eigenvectors of ${\tilde {\cal M}}$ as
\begin{equation}
{a_k} = {\cal N} \frac{1}{1 - {\tilde \lambda}_k } \sum_{i=1}^D
{\tilde \theta}_{ik} \ \ .
\label{4.7.2}
\end{equation}
Here, $\tilde{\theta}$ is the matrix that diagonalizes $\tilde{\cal
M}$ and ${\cal N}$ is chosen such that $\sum a_k^2 = 1$.

We now re--write Eq.~(\ref{4.1}) as
\begin{equation}
\sum_{k=1}^D a_{k}^2 \left( \frac{{{\tilde \lambda}}_{k}}
{1-{{\tilde \lambda}}_{k}} \right)^{\mu}
= \frac{[c \langle n \rangle]^{\mu}}{\mu + 1}  \ \ .
\label{4.2}
\end{equation}
The purpose of this re--writing is to emphasize that, in order to
reproduce the negative binomial factorial moments, it is necessary to
satisfy an infinite number of constraints, Eq.~(\ref{4.2}).  There are
only a finite number of parameters in the Hamiltonian for
finite $D$.  Specifically, there are $D(D+1)/2$ interaction parameters
to be specified \cite{footnote10}. The satisfaction of an infinite number
of constraints will evidently require the limit $D \rightarrow
\infty$.  (We shall return to this point below.)

It is interesting to note that in the $D \rightarrow \infty$ limit, we
can replace the summation in Eq.~(\ref{4.2}) by an integral in order
to obtain
\begin{equation}
\int_{-1}^{+1} d\lambda \ {a(\lambda)}^2
\left( \frac{\lambda}{1-\lambda} \right)^{\mu}
= \frac{[c \langle n \rangle]^{\mu}}{\mu + 1}
\label{4.3}
\end{equation}
where the coefficients ${a(\lambda)}^2$ now play the role of a
(non--negative) continuous weighting function.  Working with a new
variable
\begin{equation}
\eta = \frac{\lambda}{1-\lambda} \ \ ,
\label{4.4}
\end{equation}
we thus seek another weighting function, $W(\eta)$, such that
\begin{equation}
\int_{-1/2}^{+\infty} d\eta \ W(\eta) \  \eta^{\mu} =
\frac{[c \langle n \rangle]^{\mu}}{\mu + 1} \ \ .
\label{4.5}
\end{equation}
As it happens, this set of constraints allows us to determine
$W(\eta)$ by inspection. We simply obtain a step function,
\begin{equation}
W(\eta ) = \theta (c \langle n \rangle - \eta ) \theta( \eta) \ \ .
\label{4.6}
\end{equation}
This choice (barring questions of uniform convergence) is unique.  It
is readily transformed into a statement about the weighting function
${a(\lambda)}^2$.  We find ${a(\lambda)}^2 = 0$ except
\begin{equation}
{a(\lambda)}^2  =  \frac{1}{c\langle n \rangle(1 - \lambda)^2} \ \ \ \
{\rm for} \ \ 0 \le \lambda \le \frac{c\langle n \rangle}
{1+c\langle n \rangle} \ \ .
\label{4.7}
\end{equation}
This statement of the constraints will be extremely useful in finding a
scheme for picking the parameters in the Hamiltonian in order to
reproduce the negative binomial moments.

It is clear that the form of $\tilde{{\cal M}}$ as
given by Eq.~(\ref{3.1}) (for $d,d' \ne 0$) is somewhat special.
It is necessary to demonstrate
that there exists at least one scheme for picking the $D(D+1)/2$
parameters of $\tilde{{\cal M}}$ in such a way that the various
constraints for obtaining the negative binomial distribution are
satisfied.  Here, we shall simply propose two different prescriptions
and demonstrate by numerical example that these prescriptions indeed
yield the desired result.  Of course, we shall make no claims
regarding the uniqueness of our prescriptions.  The first prescription
is as follows:
\begin{quote}
Draw $D$ random numbers, $x_d$, from the interval $[0,1]$. Set
${\tilde {\cal M}}_{dd}$ equal to $x_d L$ where $L>0$.  Set all
off--diagonal elements of ${\tilde {\cal M}}$ equal to zero. For each
draw, choose $L$ such that $s_1 = c/2$.  (This sets the dispersion to
its empirical value for each draw.)
\end{quote}
The physical content of this prescription is clear. While identical
particles experience a nearest neighbor interaction which ranges from
$\approx - \mu_{\rm o}$ to $+\infty$, inequivalent particles experience a
nearest neighbor interaction in the range $-\mu_{\rm o} \ll
\epsilon_{dd'} \leq +\infty$.  The success of this prescription is
guaranteed by a comparison of Eqs.~(\ref{4.7.2}) and (\ref{4.7}).
Choosing ${\tilde {\cal M}}$ to be diagonal guarantees that the various
sums in Eq.~(\ref{4.7.2}) will be independent of $k$.  The
$\lambda$--dependence of Eq.~(\ref{4.7}) will be obtained from
Eq.~({\ref{4.7.2}) provided that the eigenvalues are distributed
uniformly over a finite, positive range.  This is ensured by our
prescription.

For our numerical studies, we consider the values of $\langle n
\rangle = 20$ and $\langle\Delta n^2 \rangle = 110$ appropriate for
the description of $p{\bar p}$ scattering at 200 GeV.  (For this
reaction, empirical values for the factorial moments are readily
available.  Qualitatively similar results are obtained for other
high--energy scattering experiments.)  These data indicate that we
should set
\begin{equation}
c \langle n \rangle = \frac{\langle \Delta n^2  \rangle - \langle n \rangle}
{\langle n \rangle} = 4.5 \ \ .
\label{4.8}
\end{equation}
We now construct the ratios
\begin{equation}
r_q = \frac{(q+1) \sum_{k=1}^D a_k^2 \left( \frac{{{\tilde
\lambda}}_{k}} {1-{{\tilde \lambda}}_{k}}\right)^{q}}{\left[c \langle
n\rangle \right]^q} \ \ .
\label{4.9}
\end{equation}
Our only constraint, $s_1=c/2$, ensures $r_1=1$.  When we have
satisfied the constraints of Eq.~(\ref{4.2}), all of these ratios
should be equal to one \cite{footnote11}.  Since there are no
parameters to adjust, this is an extremely stringent test of our
prescription.  It succeeds.  In Table~\ref{table1} we report the
results of our numerical studies for $2 \le q \le 6$ and $D = 1$, $2$,
$4$, \ldots , $512$. The notation `$\langle\langle r_q \rangle
\rangle$' is a reminder that, since our theory is now randomly drawn,
we must perform an `ensemble average' over theories as well as the
usual thermodynamic ensemble average.  For each value of $D$, this
ensemble average was performed over $10^5$ theories.

Several comments are in order.  Our prescription meets the remaining
conditions of Eq.~(\ref{4.1}) with increasing accuracy as $D
\rightarrow \infty$.  For fixed $q$, the value of $\langle\langle r_q
\rangle\rangle$ approaches $1$ like $1/D$ as $D \rightarrow \infty$.
The dispersion also vanishes (like $1/\sqrt{D}$).  Thus, as $D$
becomes large, our simple prescription converges to the results of the
NB for any fixed $q$.  For fixed $D$, the error in $\langle\langle r_q
\rangle\rangle$ and its dispersion grow as $q \rightarrow \infty$.
Our point here is that there exists at least one
simple prescription for satisfying Eq.~(\ref{4.1}) exactly.

Given the limited scope and quality of existing data, we do not need
particularly large $D$ to fit the relations between measured factorial
moments.  The value of $D = 16$ results in sufficiently small errors
and dispersions that randomly drawn dynamics have a high probability
of reproducing the relations between empirical factorial moments for
$p{\bar p}$ scattering at 200 GeV within existing experimental
uncertainties.  This value of $D=16$ is also sufficient to provide a
quantitative description of relations between factorial moments for
900 GeV $p{\bar p}$ scattering and, indeed, of all other high--energy
scattering experiments for which factorial moments are known.  To
demonstrate this, Fig.~\ref{fig1} shows the factorial moments $F_3$ to
$F_5$ for $p{\bar p}$ scattering at 200, 546 and 900 GeV as a function
of $F_2$.  We have suppressed the model--independent constant and
linear pieces of these moments, {\em i.e.}, we plot
\begin{equation}
{\tilde F}_q(F_2)=F_q(F_2)-1-\frac{q(q-1)}{2}(F_2-1) \ \ .
\label{4.10}
\end{equation}
The theoretical `bands' shown on each figure correspond to the
theoretical dispersion obtained with $D=16$.  (More than 50\% of all
randomly drawn theories lie inside this band.)  These bands were
obtained with parameters chosen to reproduce the 200 GeV data.  The
use of parameters chosen to reproduce the higher energy data would
result in somewhat narrower bands that are slightly shifted downwards.

The error bars in Fig.~\ref{fig1} represent the published
uncertainties and have roughly equal statistical and systematic
components.  Some comment regarding the statistical error is in order.
Since $F_2 (M)$ and $F_q (M)$ are drawn from the {\em same\/} data
set, significant statistical correlations can exist.  Unfortunately,
it is impossible to make a quantitative statement about such
correlations without a detailed investigation of the data reduction
techniques actually used.  We have, however, performed simulations
based on a suitably modified form of the negative binomial
distribution. The results suggest that the uncertainties shown in
Fig.~\ref{fig1} actually represent a significant overestimate of the
real uncertainties in the determination of $F_q (F_2 )$.  This belief
is also supported by the observation that the data points appear to
have a far smaller spread than their errors would indicate.  A
reduction of these errors by a factor of approximately $3$ would make
it possible to provide a convincing discrimination between the
negative binomial distribution and the Ising model distribution. Our
simulations suggest that an error reduction of this magnitude is not
out of the question.

This first prescription has the virtues of simplicity and guaranteed
success.  It can, however, be criticized on the grounds that it has
singled out the species--diagonal interactions, $\epsilon_{dd}$, for
special treatment.  We thus consider a second and more democratic
prescription in which all pairs of species are treated equivalently:
\begin{quote}
Draw the various terms $\epsilon_{dd'}$ at random over the uniform
interval
\begin{displaymath}
\epsilon_{\rm o} - \mu_{\rm o} \le \epsilon_{dd'} \le \epsilon_{\rm o}
- \mu_{\rm o} + \Delta \epsilon \ \ .
\end{displaymath}
For each choice of ${\tilde {\cal M}}$, adjust $\epsilon_{\rm o}$ in order
to reproduce the desired value of $c$.  (This can always be done.)
Adjust $\Delta \epsilon$ to reproduce the value $r_2 = 1$.  In the
event that this cannot be done, discard the draw \cite{footnote12}.
\end{quote}
This more democratic selection of ${\tilde {\cal M}}$ means that we must
actually perform the requisite diagonalizations.  This means that one
must be content with studying a smaller number of samples.  More
seriously, it is no longer elementary to establish rules which guarantee
the desired equivalence between Eqs.~(\ref{4.7.2}) and (\ref{4.7}).
It is for this reason that we have imposed the somewhat artificial
constraint that $r_2=1$.

The results of numerical studies with this second prescription are
summarized in Table~\ref{table2} for $D = 32$, $64$, $128$ and $256$.
For each value of $D$ we considered $10^{3}$ matrices.  In this case,
we plot the average values (and the corresponding dispersions) for
$\epsilon_{\rm o}$ and $\Delta \epsilon$ as well as the various ratios
$r_q$ for $3 \le q \le 6$.  The large values of $\Delta \epsilon$
required indicate that we are dealing with sparse matrices.  This has
two apparently negative consequences.  First, there will be a great
many eigenvalues close to zero.  Second, there will be only a small
preference for positive eigenvalues \cite{footnote13}. These facts
would appear to make it difficult to satisfy Eq.~(\ref{4.7}). At the
very least, we anticipate much slower convergence (as a function of
$D$) when using this second prescription.  On the other hand, the
eigenvectors associated with negative eigenvalues necessarily involve
terms of mixed signs so that there will be a greater tendency for
cancellation in the sums over ${\tilde \theta}_{ik}$ in
Eq.~(\ref{4.7.2}) for negative eigenvalues.

Somewhat surprisingly, these effects conspire to a remarkable degree.
The numerical results of Table~\ref{table2} suggest that our second
prescription also leads to the factorial moments of the negative
binomial distribution in the large $D$ limit.  This claim of success
is necessarily somewhat tentative.  The slower convergence of this
prescription suggests the need to study values of $D$ larger than
those in Table~\ref{table1}.  The need to diagonalize many large
matrices restricts us to smaller values of $D$ and smaller statistical
samples.  We have performed other tests which also suggest that this
second prescription works.  For example, we have made histograms of
the $a^{2}_{k}$ of Eq.~(\ref{4.7.2}) as a function of the variable
$\eta$ defined above.  Except for the expected `Gibbs phenomenon', we
find remarkable agreement with the distribution of Eq.~(\ref{4.6}) as
$D$ becomes large.  We have also studied the cases $D = 512$ and
$1028$ with very poor statistics.  These results are also consistent
with our assertion that the factorial moments of the negative binomial
distribution will result in the large $D$ limit.

The purpose of this second prescription has been to demonstrate the
existence of at least one democratic way to reproduce the factorial
moments of the negative binomial distribution.  More efficient schemes
may well exist.

\section{Discussions and Conclusions}
\label{sect5}

We have demonstrated that a simple extension of the one--dimensional
lattice gas model to include nearest neighbor interactions between
pairs of $D$ species of particles can reproduce the factorial moments
of the negative binomial distribution exactly provided that a number
of constraints are met.  We have further demonstrated that these
constraints can be met by simple models of the microscopic dynamics in
which all particles have the same chemical potential and in which the
nearest neighbor interactions are drawn at random according to the
prescriptions of the previous section.

Since our model approaches the negative binomial distribution as $D
\rightarrow \infty$, it will reproduce the relations between empirical
factorial moments including those obtained in $p{\bar p}$ scattering,
$e^+ e^-$ scattering and relativistic heavy ion collisions.  The only
question is whether the number of species required to provide an
adequate fit is sufficiently small to be considered `physically
reasonable'.  Given the results of Fig.~\ref{fig1} for $16$ species,
we believe that the present model is physically reasonable.  These
results also offer some understanding for both the success of cascade
calculations and the anecdotal observation that the results of
calculations are often surprisingly insensitive to the details of the
model.  According to our picture, any randomly drawn theory would be
likely to do as well (at least at the level of the factorial moments).

It has been suggested that the slopes, $\alpha_q / \alpha_2$, in plots
of $\ln{F_q}$ versus $\ln{F_2}$ can be used to distinguish between
cascade systems and systems exhibiting some critical phenomenon such
as the transition to a quark-gluon plasma \cite{Ochs}. It has been
shown that systems at the critical temperature of a second--order
phase transition display monofractal structure with $\alpha_q /
\alpha_2 \sim (q-1)$ \cite{Satz}. In contrast, cascade models lead
to a multifractal structure with $\alpha_q / \alpha_2 \sim q(q-1)$ (in
Gaussian approximation) \cite{Bialas2}. Our model (along with the
Ising model) shows that the monofractal structure of a critical
phenomenon can also be obtained from a simple but heterogeneous
(equilibrium) system. It indicates that some caution is required
before claiming insight into the nature of microscopic mechanisms on
such a basis. Indeed, the present model may be a useful test case for
proposed schemes for the detection of critical phenomena from the
study of, {\em e.g.}, factorial moments.

Before closing, we would like to offer one speculative remark
regarding the existence (or possible non--existence) of signatures
which would honestly discriminate between `interesting' critical
systems and those `uninteresting' heterogeneous systems considered
here with which they can be confused.  We suspect that the study of
global properties --- such as the factorial moments --- can only
provide convincing tests of criticality for systems whose time
evolution can be studied.  The present results suggest that it may
always be possible to devise equilibrium, heterogeneous models which
can simulate the manifestations of critical phenomena to arbitrary
accuracy in systems where the only information available is that
provided at $t = \infty$.  Evidently, this class of systems includes
all scattering experiments in nuclear and high--energy physics.  The
only way to distinguish between these classes of models is to provide
{\em independent\/} arguments that the heterogeneous alternatives to
critical systems are `too artificial' or `too unphysical' to be
accepted.  Given the present results, we suspect that this case cannot
be made.  Thus, in physics as in the rest of life, much more
information is to be gained from the study of `living' systems than
from performing autopsies on `dead' systems.

\acknowledgments

We would like to acknowledge the hospitality of NORDITA during the
summer of 1993.  We would also like to thank Nandor Balazs, Kim
Sneppen, and Peter Orland for helpful discussions.  This work was
partially supported by the U.S.  Department of Energy under grant no.
\mbox{DE-FG02-88ER 40388}.

\appendix

\section{Simplifying Factorial Moments}
\label{app1}

Consider any model for which the number operator at any site $i$ is an
idempotent:
\begin{equation}
n_i^2 = n_i \ \ .
\label{a.1}
\end{equation}
Such models include either the ordinary lattice gas model of
Sec.~\ref{sect2} or the generalized lattice gas model of
Sec.~\ref{sect3}.  Define the number operator for a single bin,
\begin{equation}
n = \sum n_i \ \ ,
\label{a.2}
\end{equation}
where $i$ runs over every site in the bin.

Now assume that we know that the following operator identity holds for
some fixed order $q$:
\begin{equation}
n(n-1)(n-2) \ldots (n-q+1) = \sum \protect\raisebox{1.2ex}{\small
$\prime$} \, n_{i_1} n_{i_2} n_{i_3} \ldots n_{i_q} \ \ .
\label{a.3}
\end{equation}
The sum extends over all sites in a given bin.  The prime denotes that
no two of the site indices, $i_1$ to $i_q$, are equal.  Note that
Eq.~(\ref{a.3}) applies at the operator level before ensemble averages
are carried out.  Our aim is to show that, if Eq.~(\ref{a.3}) applies
for $q$, it necessarily applies for $q+1$ as well.  To see this,
multiply Eq.~(\ref{a.3}) by the number operator, $n$.  This leads to
\begin{equation}
n[n(n-1)(n-2) \ldots (n-q+1)] = \sum \protect\raisebox{1.2ex}{\small
$\prime$} \, n_{i_1} n_{i_2} n_{i_3} \ldots n_{i_q} n_{i_{q+1}} + q
\sum \protect\raisebox{1.2ex}{\small $\prime$} \, n_{i_1} n_{i_2}
n_{i_3} \ldots n_{i_q} \ \ .
\label{a.4}
\end{equation}
The second term here simply corresponds to the fact that the index
$i_{q+1}$ can be equal to any one of the $q$ indices $i_1$ to $i_q$.
Using Eq.~(\ref{a.3}), we can move the final term in Eq.~(\ref{a.4})
to the left of the equation to obtain
\begin{equation}
n(n-1)(n-2) \ldots (n-q+1)(n-q) = \sum \protect\raisebox{1.2ex}
{\small $\prime$} \, n_{i_1} n_{i_2} n_{i_3} \ldots n_{i_q}
n_{i_{q+1}} \ \ .
\label{a.5}
\end{equation}
This is the desired extension of the result.  Thus, Eq.~(\ref{a.3})
will apply for all $q$ if it applies for $q=1$.  It applies trivially
for $q=1$ since the primed sum is of no meaning in this case (there
being only one term in the product) and Eq.~(\ref{a.3}) collapses to
the definition, Eq.~(\ref{a.2}).

Given the form in which the various correlation functions can be
expressed in our models, one final rearrangement of Eq.~(\ref{a.3}) is
useful.  Specifically,
\begin{equation}
n(n-1)(n-2) \ldots (n-q+1) = q! \sum_{i_1 < i_2 < \ldots < i_q}
n_{i_1} n_{i_2} n_{i_3} \ldots n_{i_q} \ \ .
\label{a.6}
\end{equation}

The final result is that Eqs.~(\ref{a.3}) and (\ref{a.6}) apply for
all $q$ for all problems in which the number operator is idempotent.
Further, one can take ensemble averages of these operator identities,
and the ensemble averages can be taken under the summations.  These
results provide a considerable technical simplification in the
evaluation of factorial moments.

\section{Special Cases of the Generalized Lattice Gas Model}
\label{app2}

Here, we wish to reduce the factorial moments of our generalized
lattice gas model to (i) the factorial moments of the ordinary lattice
gas model and (ii) the factorial moments of the negative binomial
distribution by simplifying the general expressions,
Eqs.~(\ref{3.15})--(\ref{3.17}), with the use of special assumptions
regarding the choice of the terms $s_{\mu}$.

Consider, first, the ordinary lattice gas results.  The purpose here is
an exercise in method since we already know the result to be true.
Set the various $s_{\mu}$ as
\begin{equation}
s_{\mu} = \left( \frac{c}{2} \right) ^{\mu} \ \ .
\label{b.1}
\end{equation}
We can then perform the sum in the square brackets in Eq.~(\ref{3.16})
exactly to yield
\begin{equation}
1 + \sum_{\mu=1}^{\infty} z^{\mu} s_{\mu} = \frac{1}{1-zc/2} \ \ .
\label{b.2}
\end{equation}
The desired derivative is immediately obtained as
\begin{equation}
\left[ \frac{d^k}{dz^k} \left( \frac{1}{1-zc/2} \right)^{(q-k)}
\right]_{z=0} = (q-k)(q-k+1) \ldots (q -1) \left( \frac{c}{2} \right)
^k = \frac{(q-1)!}{(q-k-1)!} \left( \frac{c}{2} \right) ^k \ \ .
\label{b.3}
\end{equation}
This immediately gives
\begin{equation}
f_k^{(q)} = \frac{q!(q-1)!}{(q-k)!k!(q-1-k)! 2^k} c^k \ \ .
\label{b.4}
\end{equation}
Comparison with Eq.~(\ref{1.4}) reveals the expected agreement.

For the case of the negative binomials, it is more convenient to start
with the generalized generating function of Eq.~(\ref{3.20}). We now
wish to set
\begin{equation}
s_{\mu} = \frac{c^{\mu}}{\mu+1} \ \ .
\label{b.5}
\end{equation}
The sum over $\mu$ can again be performed exactly to yield
\begin{equation}
1+\sum_{\mu=1}^{\infty} z^{\mu} s_{\mu} = -\frac{\ln{[1-zc]}}{zc} \ \ .
\label{b.6}
\end{equation}
Substitution of this expression into Eq.~(\ref{3.20}) yields
\begin{eqnarray}
F_q(M) & = & M^q \left[ \frac{d^q}{dz^q} \exp{\left( -
\frac{\ln{[1-zc]}} {cM} \right)} \right]_{z=0} \nonumber \\
 & = & (cM)^q \left[ \frac{d^q}{dy^q} (1-y)^{-\frac{1}{cM}}
\right]_{y=0} \nonumber \\
 & = & (cM)^q \left( \frac{1}{cM} \right) \left( \frac{1}{cM} + 1 \right)
\ldots \left( \frac{1}{cM} + [q-1] \right) \nonumber \\
 & = & (1+cM)(1+2cM) \ldots (1+[q-1]cM)
\label{b.7}
\end{eqnarray}
which is seen to be identical with the NB result of Eq.~(\ref{1.2}).

\begin{figure}
\caption{Plots of ${\tilde F}_q$ as defined in Eq.~(\protect\ref{4.10})
versus $F_2$ for $3 \le q \le 5$.  Data are taken from
Ref.~\protect\cite{UA5}. The dashed line corresponds to the
predictions of the ordinary lattice gas (Ising) model. The bold face
line represents the negative binomial distribution. The shaded area is
the prediction of the generalized lattice gas model according to our
first prescription for $D=16$.  This `band' around the NB becomes
narrower as $D$ increases.  The error bars shown are representative
for the quoted uncertainties in the data (see the text for further
comments on the error bars).}
\label{fig1}
\end{figure}

\begin{table}
\caption{The ensemble average and dispersion of $r_q$ according to
our first prescription for $2 \le q \le 6$ and various values of the
number of particle species, $D$. The average was taken over $10^5$
randomly drawn theories with $c \langle n \rangle = 4.5$ for each
draw.  The case $D=1$ corresponds to the ordinary lattice gas (Ising)
model \protect\cite{Chau} and has no dispersion.}
\label{table1}
\begin{tabular}{rr@{${}\pm{}$}lr@{${}\pm{}$}lr@{${}\pm{}$}l
                 r@{${}\pm{}$}lr@{${}\pm{}$}l}
 \multicolumn{1}{c}{$D$} &
 \multicolumn{2}{c}{$\langle\langle r_2 \rangle\rangle$} &
 \multicolumn{2}{c}{$\langle\langle r_3 \rangle\rangle$} &
 \multicolumn{2}{c}{$\langle\langle r_4 \rangle\rangle$} &
 \multicolumn{2}{c}{$\langle\langle r_5 \rangle\rangle$} &
 \multicolumn{2}{c}{$\langle\langle r_6 \rangle\rangle$} \\ \tableline
   1 & \multicolumn{2}{c}{0.75} & \multicolumn{2}{c}{0.5} &
       \multicolumn{2}{c}{0.3125} & \multicolumn{2}{c}{0.1875} &
       \multicolumn{2}{c}{0.109375} \\
   2 & 0.812 & 0.023 & 0.598 & 0.035 & 0.417 & 0.037 & 0.280 & 0.033 &
       0.184 & 0.027 \\
   4 & 0.881 & 0.046 & 0.723 & 0.078 & 0.569 & 0.091 & 0.435 & 0.091 &
       0.326 & 0.083 \\
   8 & 0.941 & 0.067 & 0.851 & 0.132 & 0.751 & 0.179 & 0.652 & 0.207 &
       0.558 & 0.218 \\
  16 & 0.981 & 0.076 & 0.951 & 0.173 & 0.916 & 0.270 & 0.879 & 0.360 &
       0.842 & 0.441 \\
  32 & 0.998 & 0.069 & 0.999 & 0.170 & 1.005 & 0.295 & 1.017 & 0.442 &
       1.039 & 0.612 \\
  64 & 1.002 & 0.050 & 1.007 & 0.128 & 1.019 & 0.229 & 1.040 & 0.355 &
       1.071 & 0.515 \\
 128 & 1.001 & 0.035 & 1.005 & 0.088 & 1.013 & 0.154 & 1.026 & 0.233 &
       1.044 & 0.327 \\
 256 & 1.001 & 0.024 & 1.003 & 0.061 & 1.007 & 0.105 & 1.013 & 0.155 &
       1.023 & 0.210 \\
 512 & 1.000 & 0.017 & 1.002 & 0.042 & 1.004 & 0.072 & 1.007 & 0.106 &
       1.012 & 0.142 \\
\end{tabular}
\end{table}

\begin{table}
\caption{The ensemble average and dispersion of $r_q$ according to our
second prescription for $3 \le q \le 6$ and various values of the
number of particle species, $D$. The average was taken over $10^3$
randomly drawn theories with $c \langle n \rangle = 4.5$ for each
draw. Also shown are average value and dispersion of $\epsilon_{\rm o}$
and $\delta \epsilon$.}
\label{table2}
\begin{tabular}{rr@{${}\pm{}$}lr@{${}\pm{}$}lr@{${}\pm{}$}l
                 r@{${}\pm{}$}lr@{${}\pm{}$}lr@{${}\pm{}$}l}
 \multicolumn{1}{c}{$D$} & \multicolumn{2}{c}{$\epsilon_{\rm o}$} &
 \multicolumn{2}{c}{$\delta \epsilon$} &
 \multicolumn{2}{c}{$\langle\langle r_3 \rangle\rangle$} &
 \multicolumn{2}{c}{$\langle\langle r_4 \rangle\rangle$} &
 \multicolumn{2}{c}{$\langle\langle r_5 \rangle\rangle$} &
 \multicolumn{2}{c}{$\langle\langle r_6 \rangle\rangle$} \\ \tableline
  32 & 0.21 & 0.37 & 114 & 75 & 0.945 & 0.025 & 0.862 & 0.059
     & 0.769 & 0.094 & 0.674 & 0.124 \\
  64 & 0.47 & 0.23 & 149 & 63 & 0.959 & 0.022 & 0.895 & 0.054
     & 0.820 & 0.089 & 0.741 & 0.122 \\
 128 & 0.62 & 0.18 & 223 & 71 & 0.973 & 0.021 & 0.930 & 0.054
     & 0.876 & 0.093 & 0.818 & 0.133 \\
 256 & 0.73 & 0.15 & 363 & 85 & 0.988 & 0.020 & 0.967 & 0.054
     & 0.941 & 0.098 & 0.910 & 0.148 \\
\end{tabular}
\end{table}


\begin{references}
\bibitem{footnote1} A reasonably successful description can be
  obtained with the replacement of $M$ by $M^{\alpha}$ in
  Eq.~(\ref{1.2}) with $\alpha$ on the order of $1/5$.
\bibitem{footnote2} While such plots are again theoretically motivated,
  they do not introduce theoretical bias into what remains a purely
  phenomenological activity.
\bibitem{footnote4} Here and in the following, we will set the inverse
  temperature, $\beta$, equal to $1$ without loss of generality.
\bibitem{footnote5} Here, the subscript $d$ means diagonal and is
  not to be confused with species $d$.
\bibitem{footnote6} This is the only real complication relative to the
  $D=1$ lattice gas model of the previous section.
\bibitem{footnote7} Again, we are taking the $N \rightarrow \infty$ limit
  for fixed $\langle n \rangle$ for which it is justified to approximate
  $1-\langle n_i \rangle$ by $1$.
\bibitem{footnote8} Given the absence of long--range order in
  one--dimensional systems, this requirement is readily satisfied.
\bibitem{footnote9} We consider only the physically interesting case where
  $|{{\tilde \lambda}_{k}}| < 1$ for all $k$.
\bibitem{footnote10} The chemical potentials have already been
  set equal to a common $\mu_{\rm o}$.
\bibitem{footnote11}  For the ordinary lattice gas, $D=1$, these ratios
  are simply $(q+1)/2^q$.
\bibitem{footnote13} Since ${\tilde {\cal M}}$ contains only positive
  elements, we are guaranteed that some preference for positive
  eigenvalues will remain.  Further, if the eigenvalue of largest
  magnitude is non--degenerate, it will be positive.
\end{references}
\end{document}